\documentclass[aps,prl,twocolumn,groupedaddress,showpacs]{revtex4}
\usepackage[latin1]{inputenc}
\usepackage[english]{babel}
\usepackage{setspace}  
\usepackage{graphicx}
\usepackage{amssymb}
\usepackage{amsmath}
\usepackage{amscd}
\usepackage{rotating}
\usepackage{color}
\usepackage{epstopdf}

\begin{document}

\title{Subband Engineering Even-Denominator Quantum Hall States}

\author{V.W. Scarola$^{1}$, C. May$^2$, M. R. Peterson$^3$, and M. Troyer$^2$  }
\affiliation{
$^1$ Department of Physics, Virginia Tech, Blacksburg, VA 24061, USA\\
$^2$Theoretische Physik, ETH Zurich, 8093 Zurich, Switzerland\\
$^3$ Condensed Matter Theory Center, Department of Physics, 
University of Maryland, College Park, MD 20742, USA
}
\date{\today}

\begin{abstract}
Proposed even-denominator fractional quantum Hall effect (FQHE) states
suggest the possibility of excitations with non-Abelian braid
statistics.  Recent experiments on wide square quantum wells 
observe even-denominator FQHE even under electrostatic tilt.  We
theoretically analyze these structures and develop a procedure to
accurately test proposed quantum Hall wavefunctions.  We find that
tilted wells favor partial subband polarization to yield Abelian
even-denominator states.  Our results show that tilting quantum wells
effectively engineers different interaction potentials allowing
exploration of a wide variety of even-denominator states.
\end{abstract}


\maketitle
 
 Experimental evidence for more than seventy odd-denominator FQHE 
 states \cite{tsui:1982} demonstrates the ubiquity of
 incompressible quantum Hall electron liquids in high mobility
 semiconductor quantum wells.  The Laughlin wavefunctions
 \cite{laughlin:1983} describe a few of these states, however, 
 the composite fermion (CF) theory includes Laughlin states into a comprehensive 
 framework that captures the
 essential physics of the observed lowest Landau level (LLL)
 FQHE states \cite{jain:1989,jain:2007}.
 
Observations of rare even-denominator FQHE states, exceptions to the
odd-denominator rule, occur in multicomponent systems \cite{suen:1992} and at 
half filling of the second LL\cite{willett:1987}, i.e., filling factor 5/2.  Spin, layer, or
subband degrees of freedom allow combinations of single-component
odd-denominator states to yield even-denominator fractions
\cite{halperin:1983}.  But at half filling of the second LL
theoretical analyses \cite{rezayi:2000} suggest that the Coulomb interaction happens to
favor the formation of a single component paired CF state, the
Pfaffian \cite{moore:1991}, a profoundly distinct state.

Proposed FQHE states possess topologically non-trivial excitations.  The common CF
states carry excitations that obey Abelian anyonic braid statistics.
Braiding two Abelian excitations changes the overall wavefunction by a
phase \cite{statistics,jain:2007}.  In contrast, excitations above the
CF paired Pfaffian state obey non-Abelian braid statistics; a braid
operates on the overall wavefunction by a matrix.  Non-Abelian excitations
have potential application in topological quantum information
processing making observation of non-Abelian excitations a key goal \cite{kitaev:2003}.

Experimental parameters (width and density) in narrow quantum well
samples allow only limited tunability.  
Recent diagonalization studies \cite{storni:2010} indicate
that in a single-layer system the Pfaffian state is likely (unlikely) to be the 
ground state for $\nu=5/2(1/2)$ within the simple Coulomb interaction model. 
By chance, including finite thickness of realistic quantum wells marginally favors 
the formation of the Pfaffian state in the second LL \cite{peterson:2008}.  A 
route towards the exploration of putative non-Abelian phases and their phase
boundaries requires construction of tunable high mobility quantum
wells designed to favor these fragile states by engineering the effective electron-electron interaction.

Recent experiments with wide square quantum wells demonstrate
even-denominator LLL FQHE in a regime with a surprisingly strong 
inter-layer tunneling \cite{luhman:2008,shabani:2009}
suggesting that explanations in terms of the usual Abelian
multicomponent states need to be revisited.  Do these observations suggest that
the new samples effectively tailor the interaction to favor a LLL
Pfaffian state \cite{papic:2009,shabani:2009}?  The competition between the (non-Abelian) Pfaffian
and (Abelian) multicomponent states remains a subtle issue
\cite{papic:2009,peterson:2010}.  Furthermore, it is 
currently unknown whether these or similar samples will favor a second LL Pfaffian or
entirely new FQHE states.   

In this Letter we show that recent experiments observing
even-denominator FQHE
\cite{luhman:2008,shabani:2009} in
wide quantum wells occur in a regime that can favor
\emph{partial} subband occupancy and therefore offer the ability to
engineer a wide array of even-denominator FQHE states.  We develop a
protocol that combines a modified high field local density
approximation (LDA) of quantum well subbands with FQHE wavefunctions
in the plane of the quantum well.  Focusing on the experimental
parameters of Ref.~\cite{shabani:2009} we argue that evidence of
even-denominator FQHE in tilted samples arises from partially
subband polarized Abelian FQHE states \cite{scarola:2001} and, in
turn, demonstrate the remarkable ability to tune among subband
balanced and imbalanced FQHE states.  Our protocol also applies to the
second LL where subband engineering can be used to tune and explore
second LL even-denominator states \cite{peterson:2010}.

We begin with a model of square quantum wells of wide width where at
most two subbands are populated.  We construct a procedure to
energetically minimize the following three dimensional model in a
strong perpendicular magnetic field, $B$:
\begin{equation}
H^{3D}=K+V_{\text{W}}+V+H_{\text{b}},
\end{equation}
where $K$ is the electron kinetic energy, $V_{\text{W}}$ is the
quantum well confinement potential, $V=\sum_{i\neq j}e^{2}/(2\epsilon|{\bf r}_i-{\bf 
r}_j|)$ is the three dimensional Coulomb interaction, $\epsilon$ is
the dielectric constant, and $H_{\text{b}}$
accounts for the energy of a plane of rigid positive background
charges.  In what follows we consider a tilted square quantum well to
model a specific set of experimental parameters \cite{shabani:2009} as
an application of our procedure: $V_{\text{W}}(z_{\perp})=\alpha
z_{\perp}+V_{0}\theta(\vert z_{\perp}\vert-w/2)$.  The extra potential, $\alpha z_{\perp}$, tilts the well.  We 
choose $V_{0}=270$ meV and a well width of $w=55$ nm. 

To determine the tilt parameter $\alpha$ we first consider the $B=0$
limit.  To solve for $\alpha$ we assume a Fermi sea in the plane and
apply usual LDA methods \cite{ortolano:1997}.  Here we assume that the
electron spins are unpolarized and use the Hedin-Lundqvist exchange
correlation energy \cite{hedin:1971} at a planar density of $\rho=1.72
\times 10^{11}$ $\text{cm}^{-2}$.  We find that the density imbalance
($7.8\times10^{10}$ $\text{cm}^{-2}$) and subband splitting measured
in Ref.~\cite{shabani:2009} ($\Delta_{01}^{B=0}=41$ $K$) are
reproduced with $\alpha=0.3$ meV/nm.

We use this value of $\alpha$ to connect to the high field experiments
of Ref.~\cite{shabani:2009} and now consider the high field, LLL limit
of $H^{3D}$.  The planar and perpendicular coordinates separate and we
assume that in the $xy$ plane the electrons are in the LLL with a
quenched kinetic energy and basis states given by: $ \phi_m(z/l_{0}) =
(z/l_{0})^{m}\exp{(-\vert z \vert^2/4l_{0}^2)}/( l_{0}\sqrt{2\pi 2^{m}
m!} ), $ where $z=x-iy$ and $l_{0}=\sqrt{\hbar c/eB}$ is the magnetic
length.  We also assume that the real spins are fully polarized (the
experiments at total filling $\nu=1/2$ in Ref.~\cite{shabani:2009} are
performed for $B\approx14T$).  We then use LDA with a spin polarized
exchange correlation energy \cite{vosko:1980} to model the electron
wavefunction perpendicular to the plane.

In wells of wide width the subband polarization,
$\gamma=(N_{0}-N_{1})/N$, is a key unknown connecting the energetics
of the plane to the energetics arising from perpendicular coordinates.
Here $N_{0}$ ($N_{1}$) denotes the number of electrons in the lowest
(first) subband with $N=N_{0}+N_{1}$.  At zero field
$\gamma$ is determined by the energetics of the $xy$-plane Fermi
surface \cite{ortolano:1997} but this is not obviously accurate in wide 
quantum wells in the LLL.

To determine the ground state we minimize the total energy per
particle,
$E^{\text{total}}_{\gamma}=E^{\perp}_{\gamma}+E^{xy}_{\gamma}$, where
$E^{\perp}_{\gamma}=(N_{0}E_{0}+N_{1}E_{1})/N$, the weighted sum of
both subband energies, arises from the quantum well confinement 
and $E^{xy}_{\gamma}$ is the correlation energy due to
the LLL planar Coulomb interaction. We calculate $E^{\text{total}}$
with the following procedure: For each value of $\gamma$ we use LDA to
compute the confinement energy,
$E^{\perp}_{\gamma}=\Delta_{\mu}^{\gamma}/2-\gamma\Delta_{01}^{\gamma}/2$,
rewritten here in terms of the $\gamma$ dependent subband energy
difference, $\Delta^\gamma_{01}$, and an effective chemical potential,
$\Delta^\gamma_\mu$.  The LDA also yields $\xi_{\sigma}$, where
$\xi_{0}$ and $\xi_{1}$ denote the lowest and first subband
wavefunctions, respectively.  The LDA output is used to construct
an effective 2D model of spin polarized LLL fermions,
$H^{\text{eff}}_{\gamma}=H^{\perp}_{\gamma}+H^{xy}_{\gamma}$, one at
each $\gamma$.  Finally, we compute $E^{xy}$ with two-component
variational FQHE wavefunctions in the subband basis at fixed $\gamma$
thus allowing global minimization of $E^{\text{total}}_{\gamma}$ in
the space of competitive FQHE wavefunctions.  Our procedure results in
a set of models, $H^{\text{eff}}_{\gamma}$, that are very sensitive to
the quantum well parameters showing that experiments can tune through
a large set of multicomponent interactions.

To model the perpendicular degrees of freedom we introduce an
effective model that captures the subband energetics:
\begin{eqnarray}
H^{\perp}_{\gamma}=\sum_{m}\left[\frac{\Delta_{01}^{\gamma}}{2}\left(n_{m,1}-n_{m,0}
  \right) -\frac{\Delta_{\mu}^{\gamma}}{2}\left(n_{m,1}+n_{m,0}\right)
  \right],\nonumber
\end{eqnarray}
where $n_{m,\sigma}=c^{\dagger}_{m \sigma}c^{\phantom{\dagger}}_{m
\sigma}$ and $ c^{\dagger}_{m \sigma}$ creates an electron in the
state $\xi_{\sigma}\phi_m$.  $H^{\perp}$ parameterizes interacting
electrons at fixed $\gamma$ in the perpendicular direction as
non-interacting fermions, with an effective chemical potential.  The
first term is the subband splitting commonly used in models of 
very wide wells 
but \emph{the second term is added to capture the energetics of
partial subband occupation in wide wells}.
\begin{figure}[t]
\includegraphics[width=3in]{fig1}
\vspace{-2.2in}
\hspace{0.16in}
\includegraphics[width=1.4in]{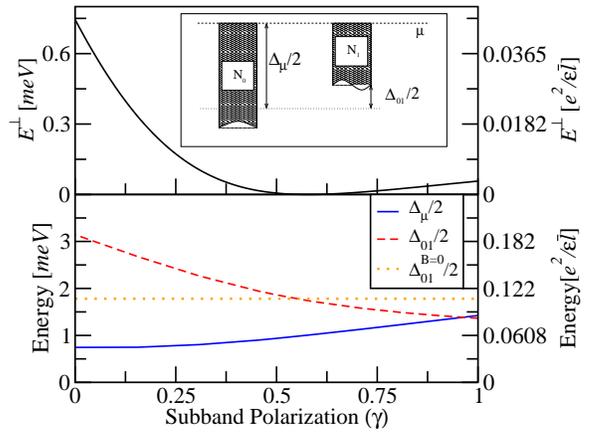}
\vspace{1.4in}
\hspace{-0.2in}
\caption{Top: Confinement energy plotted as function of subband
polarization for parameters corresponding to tilted wide well samples
of Ref.~\cite{shabani:2009}.  The right axis shows Coulomb units in
terms of the fixed magnetic length, $\overline{l}$, obtained for
$\nu=1/2$ and $\rho=1.72 \times 10^{11}\text{cm}^{-2}$.  Bottom: The
same but for the subband energy difference (dashed) and the chemical
potential offset (solid).  The dotted line plots the $B=0$ subband
energy difference.  Inset: Schematic showing partial subband occupancy
captured by an effective chemical potential.  }
\label{Evsgamma}
\end{figure}
The inset of Fig.~(\ref{Evsgamma}) shows that $\Delta_{\mu}>0$ favors
partial subband occupancy.  We obtain the very wide well limit (two fully-occupied subbands) for
$\Delta_{\mu}\gg \Delta_{01}$ and the narrow well limit (one subband occupied) with
$\Delta_{\mu}\ll \Delta_{01}$.

In the top panel of Fig.~(\ref{Evsgamma}) we see that the lowest
energy is obtained for subband polarization $\gamma\approx 0.59$.  The
large upturn near $\gamma=0$ arises because the well tilt strongly
penalizes occupancy of the second subband. The upturn near $\gamma=1$
arises from the Coulomb cost of putting all charges in the lowest
subband.  The bottom panel of Fig.~(\ref{Evsgamma}) plots
$\Delta_{\mu}^{\gamma}$ and $\Delta_{01}^{\gamma}$ versus $\gamma$ to show that they
vary in comparison to the subband splitting obtained from the zero
field calculation, $\Delta_{01}^{B=0}$.  The non-linearity
demonstrates a Stoner-like dependence on the subband-pseudospin
occupancy that arises from the competition between the kinetic energy,
well tilt, and Coulomb interaction along the direction perpendicular
to the plane.

\begin{figure}[t] 
   \includegraphics[width=3in]{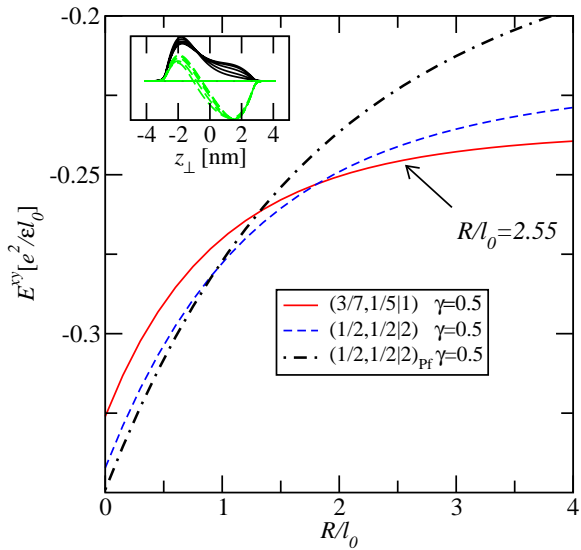}
   \caption{In-plane correlation energies of several states,
   Eqs.~(\ref{Psi}) and (\ref{PsiPf}), at $\gamma=0.5$ plotted versus
   the inter-subband interaction parameter $R$.  The arrow indicates
   the parameter relevant for the tilted well samples of
   Ref.~\cite{shabani:2009} where the $(3/7,1/5|1)$ state is shown to
   have the lowest correlation energy at $\gamma=0.5$.  Inset:
   Normalized lowest (solid) and first (dashed) subband wavefunctions
   plotted versus position in the tilted quantum well for
   several $\gamma$ between 0 and 1.  In the limit
   $\gamma\rightarrow0$ the subbands localize to mimic layer states.
   }
   \label{energyvsR}
\end{figure}

We now use output from our $\gamma$ dependent LDA calculation,
$\xi_{\sigma}$, to construct an effective interaction within the
plane.  The LLL interaction contains matrix elements of the form: $
V_{m_1m_2m_3m_4}^{\sigma_1\sigma_2\sigma_3\sigma_4}=\langle
m_{1},m_{2}\vert V_{\{\sigma\}}(r)\vert m_{3},m_{4}\rangle.  $ The
effective planar interaction is given by: $V_{\{\sigma\}}(r)=\langle
\xi_{\sigma_{1}},\xi_{\sigma_{2}}\vert V\vert
\xi_{\sigma_{3}},\xi_{\sigma_{4}}\rangle$.  The tilt breaks the
symmetry in the interaction, e.g., $V_{1010}\neq V_{0101}$.  The inset
of Fig.~\ref{energyvsR} plots the subband wavefunctions.  The tilt
localizes the subband states on either side of the well.

For the tilted wide well we find that we can ignore $V_{1100}$ and
$V_{0011}$ in the regime $0\le \gamma \lesssim 0.7$ and $\gamma=1$ to
yield a simple effective model:
\begin{eqnarray}
H^{\text{eff}}_{\gamma}\approx H^{\perp}_{\gamma}
+\frac{1}{2}\sum_{\sigma,\sigma',\{m\}}V_{\{m\}}^{\sigma\sigma'\sigma\sigma'}
c^{\dagger}_{m_1 \sigma}c^{\dagger}_{m_2
\sigma'}c^{\vphantom{\dagger}}_{m_4
\sigma'}c^{\vphantom{\dagger}}_{m_3 \sigma}.
\label{Heff}
\end{eqnarray}
The off-diagonal terms give Haldane pseudopotentials
\cite{haldane:1983} that are smaller by more than a factor of 7 than
the diagonal terms.  We assume that diagonal repulsion dominates the
inter-subband energetics and that the  omission of these small off-diagonal terms does not change the energetic ordering of trail
wavefunctions computed below using Eq.~(\ref{Heff}). 

We model the planar interaction computed with the subband
wavefunctions by a modified Zhang-Das Sarma \cite{zhang:1986} 
potential:
\begin{eqnarray}
V^{\text{MZDS}}=\left(r^{2}+d_{\{\sigma\}}^{2}/l_{0}^{2}\right)^{-1/2}
+c_{\{\sigma\}}\left(
r^{2}+W^{2}_{\{\sigma\}}/l_{0}^{2}\right)^{-1/2}\nonumber
\end{eqnarray}
where the parameters $d,c$, and $W$ are determined by fitting lowest
pseudopotentials of $V^{\text{MZDS}}$ to those of the interaction
computed using the subband wavefunctions.  We have found that
$V^{\text{MZDS}}$ gives good fits for a variety of quantum wells in
the subband basis.  For non-tilted wide wells the $c$ term must be
retained. We find largest uncertainty ($<5\%$) in the $m=1$
pseudopotential.  We have checked, by varying fitting parameters,
that $5\%$ variation in pseudopotentials does not qualitatively alter
our results.

For the tilted well sample of Ref.~\cite{shabani:2009} we focus on the
regime $0\le \gamma \lesssim 0.7$ and $\gamma=1$ and keep only the
$V_{\sigma \sigma' \sigma \sigma'}$ terms.  We then require only three
fitting parameters: $d_{0}(\gamma)\equiv d_{0000}$,
$d_{1}(\gamma)\equiv d_{1111}$, and $R(\gamma) \equiv d_{1010}$ to
approximately match all pseudopotentials.  For example, we find the
lowest $E^{\text{total}}$ to occur at: $d_{0}(0.5)=1.16 l_{0},
d_{1}(0.5)=1.37 l_{0},$ and $R(0.5)=2.55l_{0}$.

Equation (\ref{Heff}) resembles models of bilayers and very wide wells
but there are crucial differences: (i) There is a
$\Delta_{\mu}^{\gamma}$ term that favors partial subband polarization,
(ii) All terms are functions of the subband polarization, $\gamma$,
(iii) The interaction within each subband differs, e.g, $V_{1111}\neq
V_{0000}$, and (iv) In the absence of tilt we must keep off-diagonal
terms of the form $V_{\sigma \sigma \sigma' \sigma' }$.

We now construct competitive two-component variational wavefunctions
expected to minimize the energy of Eq.~(\ref{Heff}) and therefore
$H^{3D}$.  The single component LLL CF wavefunctions at filling
$n/(2pn\pm1)$, $\psi_{n/(2pn\pm1)}$, are given in the literature
\cite{jain:1989,jain:2007}.  The single component CF wavefunctions can be generalized to capture two-component states
\cite{scarola:2001}:
\begin{equation}
\Psi_{(\overline{\nu}_0,\overline{\nu}_1|m)}= \prod_{r,j}(z_j-w_r)^m
\psi_{\overline{\nu}_0}[\{z_k\}] \psi_{\overline{\nu}_1}[\{w_s\}],
\label{Psi}
\end{equation}
where the fully antisymmetric wave function $\psi_{\overline{\nu}}$ is
a single component state at filling factor $\overline{\nu}$.
Halperin's wave functions \cite{halperin:1983} are obtained as special
cases for $\overline{\nu}_0=1/m'$ and $\overline{\nu}_1=1/m''$.  In
order to ensure that the electrons of each component occupy the same
area, $N_0$ and $N_1$ must be related by $ N_0 \overline{\nu}_0^{-1}+m
N_1 = N_1 \overline{\nu}_1^{-1} + m N_0, $ thus
$\nu=(\overline{\nu}_0^{-1}+mN_{1}/N_{0})^{-1}+(\overline{\nu}_1^{-1}+mN_{0}/N_{1})^{-1}$.
Some of the $\gamma=0$ states of Eq.~(\ref{Psi}) have been shown to be
energetically competitive with favorable ground states at several
different filling factors in bilayer systems (small tunneling and
equal subband population)
\cite{scarola:2001}. Table~\ref{table_wavefunctions} lists a larger
set.  The constituent wavefunctions, $\psi_{\overline{\nu}}$, with the
lowest energy gap yield an upper-bound for the energy gap of each multicomponent
state.  For example, the gap of the $(3/7,1/5|1)$ state is no larger
than that of the $\psi_{1/5}$ state while $(1/2,1/2|2)$ is a gapless
CF Fermi sea \cite{jain:2007}.  We also include the multicomponent
Pfaffian states in our comparison:
\begin{equation}
\Psi_{(\overline{\nu}_0,\overline{\nu}_1|m)_{\text{Pf}}}=
\prod_{r,j}(z_j-w_r)^m \psi_{\text{Pf}}[\{z_k\}]
\psi_{\text{Pf}}[\{w_s\}],
\label{PsiPf}
\end{equation}
where $\psi_{\text{Pf}}$ is the Pfaffian wavefunction
\cite{moore:1991}.

We compute the ground state energies of several candidate
wavefunctions in the subband basis at $\nu=1/2$ for the parameters of
Ref.~\cite{shabani:2009} using variational Monte Carlo.  We use the
spherical geometry \cite{jain:2007} to compute energies in finite size
systems and extrapolate our results to the thermodynamic limit by 
adding the background energy: $E_{\text{b}}=-\left[ N_{1}^{2}F(d_{1})+ N_{0}^{2}F(d_{0})+
2N_{0}N_{1}F(R)\right]/2$, where $F(x)\equiv (e^{2}/ 2\epsilon
l_{0})(-x/R_{s}+\sqrt{4+x^{2}/R_{s}^{2}})$ and $R_{s}$ is the radius
of the sphere in units of the magnetic length.

Fig.~\ref{energyvsR} shows a representative comparison of $E^{xy}$ for
three of the lowest energy states at $\gamma=0.5$.  Here the parameter
$R$ is varied to test the robustness of the lowest energy state.  The
Monte Carlo error bars are smaller than the line width.  The lowest
energy state at $\gamma=0.5$ is found to be the incompressible
$(3/7,1/5|1)$ state.  The compressible Fermi sea state, $(1/2,1/2|2)$,
is nearby in energy.  The Pfaffian state becomes lower in energy for
the unphysical regime of $R<l_{o}$.

\begin{table}
\begin{tabular}{|c|c|c|c|c|} 
\hline
  $(\overline{\nu}_0, \overline{\nu}_1 | m )$ & $n_0$ & $n_1$ &$\gamma$ \\ \hline \hline
   $(1/3,1/3 |1)$  & 1 & 1& $ 0$\\ 
  $(3/7,1/5|1)$ & 3 & 1 & 1/2\\ 
  $(5/11,1/7|1)$ & 5 & 1 & $2/3$    \\
  $(1/2,1/2|2)$ & $\infty$ & $\infty$ & any   \\
  $(1/4,1/4|0)$ & $\infty$ & $\infty$ & 0    \\ \hline
\end{tabular}
\caption{Several possible states, Eq.~(\ref{Psi}), at half filling.
The last column, polarization, shows that two of the incompressible
states are locked at partially polarized configurations.
\label{table_wavefunctions}}
\end{table}

We have compared wavefunctions at several subband polarizations to
globally minimize the total energy.  We expect partially subband
polarized states to be favored by the competition between the Coulomb
interaction along the direction perpendicular to the plane and the
quantum well tilt.  We find that the small gain in $E^{\perp}$ favors
the $(3/7,1/5|1)$ state over all others in a parameter window
corresponding to the parameters relevant for the tilted well
experiments of Ref.~\cite{shabani:2009}, $\alpha=0.3$ meV/nm, see
Table~\ref{table_energies}.  We also conclude that small changes in
$\alpha$ will favor the partially subband polarized compressible
state, $(1/2,1/2|2)$.

\begin{table}
\begin{tabular}{|c|c|c|c|c|} 
\hline
$(\overline{\nu}_1, \overline{\nu}_2 | m )$ &$\gamma$ & $E^{xy} [e^{2}/\epsilon l_{0}]$ & $E^{\perp} [e^{2}/\epsilon \overline{l}]$& $E^{\text{total}} [e^{2}/\epsilon \overline{l}]$ \\ \hline \hline
 $(1/3,1/3 |1)$  &  0& -0.24858(2)& 0.045 & -0.20358(2)\\ 
 ${\bf (3/7,1/5|1)}$ & {\bf 1/2}& {\bf -0.24542(4)}& {\bf 0.00036}& {\bf -0.24506(4)}\\ 
   $(1/2,1/2|2)$ & 1/2 & -0.2440(4) & 0.00036 &  -0.2436(4) \\
    $(1/2,1/2|2)_{\text{Pf}}$ &1/2 & -0.2214(4) & 0.00036 & -0.2210(4) \\
  $(1/2,0|0)$ &  1 & -0.246394(4)& 0.0035  &  -0.242894(4)\\
 $(1/2,0|0)_{\text{Pf}}$ & 1 & -0.24338(5)& 0.0035 & -0.23988(5) \\
 \hline
\end{tabular}
\caption{Comparison of planar correlation energies (third column) and
confinement energies (fourth column) for several $\nu=1/2$ states.
The energies were computed for parameters relevant for the tilted well
sample of Ref.~\cite{shabani:2009}.  The lowest total energy is found
for the incompressible $(3/7,1/5|1)$ state.
\label{table_energies}}
\end{table}

Our procedure provides the following physical picture of the
experiments in Ref.~\cite{shabani:2009} at $\nu=1/2$.  A symmetric
wide well favors the compressible $(1/2,1/2|2)$ state with $\gamma \approx 0$ (or
the incompressible  $\gamma=0$ $(1/3,1/3|1)$ state depending on $\rho$ and $w$).
As the density is made more asymmetric via a tilt the CF Fermi sea state 
continuously depopulates its higher 
subband (it is a subband-paramagnet).  Near $\alpha\approx 0.3$ meV/nm each subband becomes
localized on either side of the well to approximate layer-like behavior 
but with asymmetric parameters in the subband
basis.  The incompressible asymmetric state, $(3/7,1/5|1)$, is then
favored in a narrow parameter window.  We find that this state is
responsible for the indications of FQHE observed under tilt in
Ref.~\cite{shabani:2009}.  As the state is made more asymmetric the
higher subband state becomes further depopulated to again favor
$(1/2,1/2|2)$.  Our analysis shows that the samples of 
Ref.~\cite{shabani:2009} allow tuning among partially subband polarized
multicomponent states.

We have shown that quantum wells with two active subbands can exhibit
a wide variety of partially subband polarized, even-denominator FQHE
states.  Our procedure can be used to find parameter regimes where
incompressible partially subband polarized FQHE states arise as a
function of sample density, width, and tilt (e.g., a mixed paired
state \cite{scarola:2002}).  Eq.~(\ref{Psi}) also provides candidates
for recently observed \cite{luhman:2008,shabani:2009} LLL
$\nu=1/4$ features (e.g., $(3/13,1/7|3)$), and the second LL FQHE
where tilt can be used to tune between a variety of subband
polarizations thereby allowing one to engineer a larger class of
interaction potentials.  Furthermore, Eq.~(\ref{Psi}) offers several
candidate partially polarized real-spin FQHE states.

We thank S. Das Sarma, J.K. Jain, and M. Shayegan for helpful discussions.  Work at University of Maryland (MRP) was supported by Microsoft Q.

\end{document}